\RequirePackage[2020-02-02]{latexrelease}
\documentclass[twocolumn,showpacs,preprintnumbers,amsmath,amssymb]{revtex4}

\usepackage{graphicx}

\usepackage{amssymb}
\usepackage{graphicx}
\usepackage{amsmath}

 \usepackage[utf8]{inputenc}

\usepackage{graphicx}
\usepackage{dcolumn}
\usepackage{bm}

\begin{document} 
\title{Background gravitational waves as signature of the adiabatic expansion of a black body that represents the dark universe}
\author{Cl\'audio Nassif Cruz} 
\altaffiliation{UFOP: Universidade Federal de Ouro Preto, Morro do Cruzeiro, Bauxita, 35.400-000-Ouro Preto-MG, Brazil. \\
email: claudionassif@yahoo.com.br} 
 \author{A. C. Amaro de Faria Jr.}
 \altaffiliation{UTFPR-GP: Federal Technological University of Paran\'a, Av. L. Bastos, 85053-525, Guarapuava-PR, Brazil.\\
email: atoni.carlos@gmail.com}


\begin{abstract}
We propose a toy model of a spherical universe made up of an exotic dark gas with temperature $T$ in thermal equilibrium with a black-body in adiabatic 
expansion. Each particle of this exotic gas mimics a kind of particle of dark energy represented by the vacuum energy, being quantized into virtual particles with extremely small masses that form such gas representng the own tissue of the expanding space-time governed by a negative pressure whose
origin is the equation of state (EOS) of vacuum, i.e., $p=-\rho$, where 
$\rho$ is the vacuum energy density. So, each vacuum particle occupies a tiny area of space so-called Planck area $L_p^{2}$, which represents the minimum area of the whole space-time given by the spherical surface with area $4\pi R_H^2$, where $R_H$ is the Hubble radius. We realize that such spherical surface is the surface of the black-body for representing the dark universe as if it were the surface of an expanding balloon. Thus, we are able to derive the law of universal gravitation, thus leading us to understand the cosmological anti-gravity. We estimate the very small order of magnitude of the cosmological constant and the acceleration of expansion of the dark sphere. In this toy model, as the dark universe can be thought of as a large black body, when we obtain its power and frequency of emission of radiation, we find very low values. We conclude that such radiation and frequency of the black body made up of dark energy is a background gravitational wave with very low frequency in the order of $10^{-17}$Hz due to the slight stretching of the fabric of space-time.
\end{abstract}

\maketitle

\section{Introduction} 

In the early 2000s a series of works flourished combining General Relativity, Scalar Fields and extra dimensions
\cite{Maartens,Arkani,Langlois,Goldberger,DeWolfe,Bazeia}. All these works are based on Action $S$ in $D$ dimensions whose structure is composed of curvature, scalar fields and corresponding momentum energy tensor, i.e., we have the action
$S =\int d^{D}x\sqrt{|g|}\left[\frac{1}{2} \partial_{\mu}\phi \partial^{\nu}\phi -\frac{1}{2}m^2\phi^2 -V(\phi) \right]$, 

It is known that the dark energy acts as a tenuous uniform distribution of energy and with property of having a negative pressure, governed by the EOS of vacuum, namely $p=-\rho$, thus leading to the accelerated expansion of the universe. This is the reason why we will build a toy model for an expanding dark gas made up of $N$ virtual particles for representing the dark energy, i.e., we admit that the vacuum energy is composed of $N$ virtual particles working like a gas that expands, since it is governed by the EOS $p=-\rho$, where $p$ is its negative pressure and $\rho$ is its energy density. 

Although the true nature of the dark energy is unknown, there are some models that seek to understand its nature. Simpler candidates include the cosmological term (cosmological constant) introduced by Einstein to modify the field equations of the General Theory of Relativity, or a scalar field that is extremely weak and has properties that are consistent with repulsive macroscopic behavior.

Actually, the present toy model predicts a cosmological parameter, which depends on the size (radius $r$) of the sphere (universe), i.e., $\Lambda(r)\propto 1/r^3$ (Eq.4), instead of simply a cosmological constant that does not depend on radius $r=ct$, where $t$ is the time. Of course the present toy model will show that the well-known cosmological constant $\Lambda_0$ 
is obtained just for the current universe, i.e., only for the Hubble radius $r=R_H$, where we have $\Lambda_0=A/R_H^3$ (Eq.4), where $A$ is a constant to be determined. 

In short, this means that the toy model predicts a cosmological parameter $\Lambda(r)$ that varies with radius and time instead of simply being constant, so that it could be thougth of as a {\it cosmological scalar field} that deacreases with the increasing of the radius and time of the universe. Therefore $\Lambda(r)$ is not constant throughout the history of the universe, 
being a good candidate for describing the evolution of dark energy density, while the cosmological constant model ($\Lambda_0$) is usual only for the radius $r\sim R_H$ (Eq.4), i.e., for the current universe.  

Therefore, based on this idea of a cosmological scalar field $\Lambda(r)$ predicted by the present toy model of a spherical universe, the idea of a cosmological constant $\Lambda_0$ 
that does not depend on the age and radius of the universe must be abandoned and replaced by a variable parameter $\Lambda$. 

The universe is approximately $13.7$ billion years old with Hubble radius 
$R_H\sim 10^{26}$m. It is composed of about $4\%$ ordinary matter (such as hydrogen, helium, etc.), $23\%$ dark matter, which gravitationally attracts, and $73\%$ dark energy, whose nature is mysterious and acts as a constant energy density associated with the vacuum energy and with negative pressure, working like a cosmological anti-gravity, i.e., with EOS $p=-\rho$, which 
describes the expansive nature of the dark gas that represents the vacuum 
energy filling out the space-time tissue.  

The toy model treated in this paper allows us to make a connection between Unruh effect and gravity, which can be extended to the cosmological anti-gravity in view of the fact that we consider a black body made of dark energy. This issue will be explored in the section 2. 

The thermodynamical aspect of gravitation is also another puzzle that should be deeper explored. Recent investigations have shown that gravitational field equations in a wide class of models can be interpreted with a thermodynamical origin. This idea, originally due to Sakharov\cite{1}, has different forms of implementation\cite{1,2,3,4,5,6,7,8,9,10,11,12,13,14,15,16} and also other approaches\cite{17,18,19}.

In the section 3, we use the toy model of an expanding dark sphere given in the section 2 for representing the current universe with $r=R_H$. Due to the Unruh effect treated in this frontier $R_H$, we will be able to obtain a very low thermal energy $K_BT$ whose associated frequency is of an order of magnitude far below the Cosmic Microwave Background (CMB). Due to the small acceleration of the universe, such a very low frequency ($\approx 10^{-17}$Hz) will be interpreted as being the frequency of a gravitational wave emitted by the dark energy due to the expansion of the fabric of space-time. 

In the section 4, as the present toy model deals with the dark universe as being the spherical surface in Fig.1 with an extremely low temperature, we will apply the Planck law for black body in order to obtain the power of emission of the background radiation with a frequency in the order of $10^{-17}$Hz. Actually, as such frequency is far outside the measurable electromagnetic spectrum and it is also a result of the expansion acceleration of a dark mass (no electric charge), so it has gravitational origin. It is
due to the Unruh effect, namely the small expansion acceleration of the dark energy on the surface of the sphere is connected to a very low temperature. 

Dark energy does not emit any kind of electromagnetic wave. Hence the fact that such energy is dark, but it has gravitational waves signature, whose
origin is the radial accelerated expansion of the sphere of 
dark gas (dark energy), i.e., such accelerated expansion of the dark 
energy emits gravitational waves, which comes from all directions of space,
thus generating the background gravitational waves. 

In the section 5, the toy model predicts a variable cosmological ``constant'' so-called
the cosmological scalar field $\Lambda(r)$ as alternative to the cosmological constant
$\Lambda_0$. The scalar field $\Lambda(r)$ is able to incorporate the background gravitational waves that control the dynamics of expansion of the universe, so that the concept of a cosmological constant, which does not vary with the radius and time of the universe should be abandoned.

\section{Unruh effect and anti-gravity}  
 
In the early universe, space-time was confined into the Planck area $L_p^2(\sim 10^{-70}m^2)$. This area has expanded very quickly as described by cosmic inflation. After inflation, it continued to expand much slower until reaching the current area $4\pi R_H^2$, $R_H(\sim 10^{26}m)$ being the Hubble radius. 
 
The present toy model uses a dimensional compaction that preserves some 
physical properties like the universal gravitation law. This means that the expanding universe $3D$ works like a spherical surface $2D$ with area
$4\pi R_H^2$. The reason for such model of compactation is that it is 
considering a spherical surface $2D$ for representing the fabric of space-time $4D$. 
 
In this toy model, let us consider the universe made up of an exotic gas of $N$ particles in a thermal equilibrium with the particles on the surface of the cavity of a spherical black-body, which represents a background frame for the fabric of space-time. 
 
 As the cavity $2D$ of the black-body represents the space-time, we should consider that each particle of the exotic gas is inside the tiny Planck area ($L_P^2\sim 10^{-70}m^2$) of the big dark cavity with area $4\pi R_H^2$. So we conclude that the spherical surface with radius $R_H$ can be divided in $N$ areas of Planck with $N$ exotic dark particles. So we obtain $N=4\pi R_H^2/L_P^2$, where $N$ is the number of dark particles over this spherical surface, $R_H\sim 10^{26}$m being the Hubble radius with about $13.7$ Gyears.
 
We estimate the large number $N$ of exotic particles of dark energy inside
each Planck area of the spherical surface that represents the tissue of the 
space-time, namely $N\sim 10^{122}$ (Fig.1).   

We know that the Planck length $L_p$ is given by $L_P=\sqrt{G\hbar/c^3}$.
 
Each particle of such an exotic dark gas over the spherical surface is expanding radially in only one direction $r$. So its total thermal energy is $E=Mc^2=\frac{1}{2}NK_BT$, where $E=E_{dark}$, with $N\sim 10^{122}$. Such equation represents the total dark energy of a dark exotic gas made up of about $10^{122}$ dark particles that governs the accelerated expansion of the universe. 
 
By using $N=4\pi R_H^2/L_P^2$, $L_P=\sqrt{G\hbar/c^3}$, we rewrite the thermal energy of the gas as follows: 

\begin{equation}
 E=Mc^2=\left(\frac{2\pi c^3 K_B}{G\hbar}\right)TR_H^2, 
\end{equation}
where $R_H$ is the Hubble radius and $E=Mc^2=E_{dark}=M_{dark}c^2$. 

We should stress that such an exotic gas is not a true gas with its particles moving randomly over a $2$-dimensional surface. Actually, this exotic gas is a mimic of a gas, as we have considered only one degree of freedom represented by the radial direction $r$ of motion of the 
$N$ particles.

\begin{figure}
\includegraphics[scale=1.0]{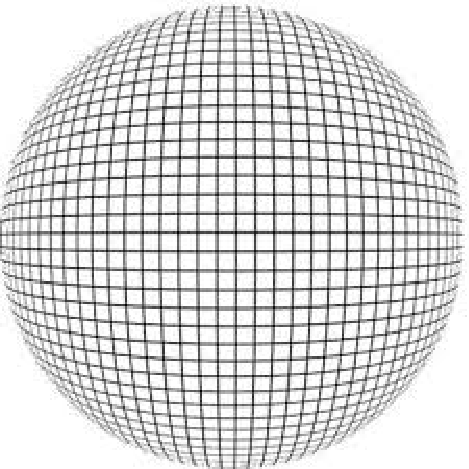} 
\caption{The surface of the dark sphere has a number $N\sim 10^{122}$ of Planck areas $L_P^2$. Each one of them is connected to a dark infinitesimal energy as being a dark particle of vacuum. The total mass $M$ in Eq.(1) is the total dark mass $M_{dark}$. A baryonic proof particle over the surface of the sphere experiences an acceleration based on the Unruh effect. As the sphere is made up of dark energy, we have an anti-gravitational acceleration associated with the cosmological constant given for $r=R_H$. Thus such a proof particle of matter is pushed out of the dark sphere (universe) due to the anti-gravity.} 
\end{figure} 

The proof particle of matter is over the spherical surface. Such particle is in thermal equilibrium with the sphere. So, as this particle has a radial acceleration ($a(r)$), the Unruh effect\cite{20} states that such a particle experiences a thermal bath with the same temperature $T$ of the sphere, i.e., $T=\hbar a/2\pi K_B c$. 

Due to the thermal equilibrium between the accelerated proof particle and the spherical surface with temperature $T$, we use the Unruh effect in Eq.(1), so that we obtain 

\begin{equation}
 E=M_{dark}c^2=\left(\frac{2\pi c^3 K_B}{G\hbar}\right)\left(\frac{\hbar a}{2\pi K_B c}\right)R_H^2, 
\end{equation}
from where we get 

\begin{equation}
a=\frac{GM_{dark}}{R_H^2}.
\end{equation}

Eq.(3) is the law of universal gravitation. It was obtained within a scenario of quantum-gravity (dark energy), i.e., we have a dark mass $M_{dark}$ made up of an
exotic gas that fills the whole space-time (Fig.1). This gas is expansive, which leads to the accelerated expansion of the spherical surface (fabric of the space-time), working like an anti-gravitational acceleration.

In a recent paper\cite{Lambda}, as it was shown from first principles that the equation of state of vacuum (EOS) shows an anti-gravity due the negative pressure $p=-\rho$, $\rho$ being the vacumm energy density associated with a preferred reference frame of an invariant minimum speed $V$\cite{Lambda} of a modified relativity\cite{Lambda}, we are led to conclude that the 
Eq.(3) represents the universal gravitation law applied to the accelerated expansion of the universe by avoiding its collapse. In view of this, we should interpret Eq.(3) working like a cosmological anti-gravity. So, let us just rewrite Eq.(3) as $a_0=-\frac{GM_{dark}}{R_H^2}$, where the negative sign ``$(-)$'' shows that the acceleration is repulsive for a given proof particle of matter on the surface of the dark sphere. $R_H$ is the Hubble radius and $M_{dark}$ is the dark mass associated with the dark energy $E_{dark}=M_{dark}c^2$. This represents $73\%$ of the whole universe. 

According to recent observations\cite{21}, the density of matter in the universe is about $3\times 10^{-30}g/cm^3$. Such density includes the contribution of dark matter plus the density of luminous matter as stars and galaxies. This represents about $27\%$ of the whole universe, where $23\%$ is the dark matter and $4\%$ is the luminous matter. 

The visible universe has about $13.7$ Gyears. By considering the density associated with luminous and dark matter, which represents $27\%$ of the whole universe, we find a mass of about $3\times 10^{55}$g. However, as the dark energy is $73\%$ of the universe, we compute $M_{dark}$ about $8.1\times 10^{55}$g. This leads us to conclude that the entire energy content can be estimated as being the own dark energy $M_{dark}$ of the order of $10^{56}$g. Thus, by making $M_{dark}\sim 10^{53}$Kg, $R_H\sim 10^{26}$m and $G\sim 10^{-10}Nm^2/kg^2$ in the law of universal gravitation for representing anti-gravity, we obtain the acceleration of anti-gravity for a proof particle on the surface of the dark sphere, namely $a_0\sim -10^{-9}m/s^2$. Such result is expected as the current acceleration of expansion of the universe (anti-gravity) is very low. 

We must stress the fact that the equation for obtaining the negative acceleration $a_0$ of a proof particle on the surface of the sphere of radius $R_H$ is the Newtonian law of gravitation. This is a good approximation once the dark mass in the order of $10^{56}g$ is uniformly distributed in a big sphere with volume in the order of $10^{78}m^3$, giving a too low density, i.e., we have $\rho_{dark}\approx 8.1\times 10^{55}g/(4/3)\pi 10^{78}m^3\approx 10^{-23}g/m^3$ or $\rho_{dark}\sim 10^{-29}g/cm^3$. Such very low density of the dark sphere is in good agreement with observations. So it justifies the use of the Newtonian approximation for a weak anti-gravity. However, if this mass in the order of $10^{53}Kg$ has a too small volume, so that the sphere presents a high density, so this result cannot be used for the approximation of classical gravitation. However, as we are only interested in the current state of the universe with an extremely low density of vacuum energy ($\sim 10^{-29}g/cm^3$), the Newtonian approach works well for obtaining the tiny value of the cosmological constant  $\Lambda_0$ given by the derivative of acceleration for $r=R_H$, namely $\Lambda_0=da/dr|r=R_H$. 

The cosmological constant $\Lambda$ is the acceleration rate per unit of distance $r$ ($s^{-2}$ in SI units). Thus we have $\Lambda=da/dr$. So, for $r=R_H$, we get its current value $\Lambda_0$, namely: 

\begin{equation}
\Lambda_0=\left[\frac{da}{dr}\right]_{r=R_H}=\frac{2GM_{dark}}{R_H^3}\sim
10^{-35}s^{-2},    
\end{equation}
where $a=-GM_{dark}/r^2$ is repulsive, being $M_{dark}\sim 10^{53}$Kg. 
As $\Lambda_0>0$, we have an anti-gravity as expected. 

We have obtained $\Lambda_0=\Lambda(R_H)\sim 10^{-35}s^{-2}$. This is the order of magnitude of the cosmological constant $\Lambda_0$ in agreement with the current observational data\cite{22,23,24,25,26,27,28}. 

Alternatively, in the natural unity system, we write $\Lambda_0=\left[\frac{1}{c^2}\frac{da}{dr}\right]_{r=R_H}=\frac{2GM_{dark}}{c^2R_H^3}\sim 10^{-35}s^{-2}/10^{17}m^2s^{-2}\sim 10^{-52}m^{-2}$. 

Here it is important to call attention to the fact that there are two equivalent unity systems to express the cosmological constant, namely the natural unity system, i.e., the unity of $\Lambda$ is $U(\Lambda)=m^{-2}$ and the international unity system, i.e., the unity of $\Lambda$ is $U(\Lambda)=s^{-2}$. We must stress that both unity systems are in fact equivalent.

Knowing that the current vacuum energy is $\rho_0\sim 10^{-29}g/cm^3$, so we
find $\Lambda_0=\frac{8\pi G}{c^2}\rho_0\sim 10^{-35}s^{-2}$, which provides 
the international unity system for representing the tiny value of 
$\Lambda_0$. On the other hand, we can alternatively write $\Lambda_0=\frac{8\pi G}{c^4}\rho_0\sim 10^{-52}m^{-2}$, which provides the natural unity system for representing the same cosmological constant $\Lambda_0$.
The constant $k=8\pi G/c^4$ or $k=8\pi G/c^2$ appears as proportionality
constant in Einstein equation given in the natural unity system or 
in the international unity system respectively, where we write Einstein equation with the presence of the cosmological constant as being 
$R_{\mu\nu}-\frac{1}{2}Rg_{\mu\nu}+\Lambda g_{\mu\nu}=kT_{\mu\nu}$. 

In this paper, we just opted for the international unity system as shown in Eq.(4).%
We must be aware of the fact that the very high
values obtained for the cosmological constant and the
vacuum energy density by using the Quantum Field Theory for describing the quantum vacuum energy have a high discrepancy of about 120 orders of magnitude beyond their observational values. This puzzle is well known as the \cite{29}\cite{30}, and it is deeply investigated in a modified relativity with an invariant minimum speed $V$ by representing vacuum in such a modified space-time with the presence of the cosmological constant\cite{Lambda}. 

\section{Gravitational waves as a signature of the dark energy due to accelerated cosmic expansion} 

According to the present toy model, a proof particle of matter over the dark spherical surface experiences a thermal equilibrium with the sphere with a temperature $T$, which is proportional to its radial acceleration ($a=a(r)$), i.e., $T\propto a$. This is the so-called Unruh effect, from where we can obtain the very low thermal energy, namely: 

\begin{equation}
K_BT=\frac{\hbar a}{2\pi c}=\frac{G\hbar M_{dark}}
{2\pi cR_H^2}\sim 10^{-51}J,
\end{equation}
where $a\sim 10^{-9}m/s^2$ for $R_H\sim 10^{26}m$, $M_{dark}\sim 10^{53}$Kg and $T\sim 10^{-28}$K. 

\subsection{The very low thermal energy from the dark energy far below the electromagnetic range as a signature of gravitational waves} 

There is a very low thermal energy associated with the dark energy. Thus, as we will show that such energy is too low, in fact it emits a very low frequency. 
 
From Eq.(5), we see that
 
 \begin{equation}
  K_BT\sim 10^{-51}J. 
 \end{equation}
 
As $K_B\sim 10^{-23}J/K$, we find $T\sim 10^{-28}$K. Such temperature is too low, so that the predominant frequency emitted by the black body of dark energy (Fig.1) is 
 
 \begin{equation}
 \nu=\frac{K_BT}{h}=\frac{a}{4\pi^2 c}=\frac{G M_{dark}}
{4\pi^2cR_H^2}\sim 10^{-17}Hz, 
 \end{equation}
 where $a=GM_{dark}/R_H^2\sim 10^{-9}m/s$. We have used $h\nu=K_BT$ and Eq.(5), where $\hbar=h/2\pi$. 
 
 Here it is important to stress that the visible universe is also treated as a model of black body, since it is permeated by a cosmic microwave background (CMB) radiation with a temperature of about $2.725$K associated with a peak frequency of $160.4$GHz, which corresponds to a wavelength of $1.9$ mm. Such electromagnetic radiation is practically isotropic with very small fluctuations and comes from all directions of space. 
 
 The present model of dark sphere (dark universe) given by a black body is analogous to the first one; however it emits an isotropic background radiation of gravitational origin. Its frequency is about $10^{-17}$Hz, being far below the measurable electromagnetic frequency. Actually, it is not an electromagnetic wave, being a background gravitational wave emitted by the dark energy permeating the whole space, as the dark energy leads to an accelerated expansion of the fabric of space-time represented by the surface of the dark sphere (Fig.1). 
 
 We realize that the background gravitational wave emitted by the expanding dark energy with an extremely low frequency $\nu_g\sim 10^{-17}Hz$ has a cosmological wavelength $\lambda_g=c/\nu_g\sim 10^{25}m$, which represents a supergiant background gravitational wave. It has a very weak signal and it is the remnant of the cosmic inflation in the current universe. 
 
 Here it must be stressed that the background gravitational waves are formed in the whole fabric of space-time represented by the spherical surface in accelerated expansion, therefore coming from all directions of space, since there is no local source of emission of gravitational waves from the cosmological vacuum, but it is the space-time itself as a whole 
 that generates a background isotropic radiation of gravitational origin, i.e., a background cosmic gravitational radiation. Such a background cosmic gravitational waves arises because of
 the stretching of the tissue of space-time due to the cosmological anti-gravity governed 
 by the EOS $p=-\rho$ that leads to the accelerated expansion of the spherical surface in
 this toy model. Thus we conclude that the accelerated expansion of the fabric of space-time
 given by the spherical surface like the surface of an inflatable balloon causes the emission of background gravitational waves with frequencies in the order of $10^{-17}$Hz. 
 
 Although the background gravitational waves are just consequence of the accelerated cosmic 
 expansion of the fabric of space-time (cosmological anti-gravity), it seems that such waves themselves do not generate anti-gravity effect, at least in a first approximation, since the gravitational waves generated by the cosmological anti-gravity are very weak. However, if we admit background gravitational waves with much higher frequencies in the much younger universe with radius $r<<R_H$ and $\Lambda>>\Lambda_0$, when the acceleration of expansion was much larger, thus, in a second approximation, it is possible that the background gravitational waves generated a most significant anti-gravity effect and so on in a self-feeding process between anti-gravity effect and gravitational waves that reaches higher orders as $\Lambda$ increases. But, as the present model deals only with the current universe, having a tiny value of 
 $\Lambda(\sim 10^{-35}s^{-2}$) and a very low frequency for the background gravitational wave $\sim 10^{-17}$Hz, all possible effects of higher orders of self-feeding process between anti-gravity effect and gravitational waves can be neglect, which does not prevent future explorations towards generalized models where strong gravity is taken into account.

 The CMB cloaks the effects of the dark energy. What this means is that if we just assume the absence of CMB, the background temperature would be much lower in the order of $10^{-28}$K. Although this is not true due to the predominance of CMB, even so the present model predicts a gravitational wave of frequency approximately $10^{-17}$Hz emitted by the dark energy permeating isotropically the whole expanding universe. This lowest frequency of gravitation origin is the fossil of the inflationary phase given in the current universe. 
 
 A curved signature in the cosmic microwave background (CMB) provides proof of inflation and space-time ripples. A CMB imprint from primordial gravitational waves would be indirect evidence, like the high-water mark left by a receding tide, but what of detecting the waves themselves? At present, the instruments most sensitive to locally produced gravitational waves are interferometers such as the dual Laser Interferometer Gravitational Wave Observatory (LIGO) installations. Each sends laser light down four-kilometer arms to look for interference effects that would be caused by passing gravitational waves stretching space in one direction and compressing it in the other. In future decades, space-based interferometers could use the same principles on much larger scales, bouncing laser light across the solar system, to directly detect the fainter primordial gravitational waves like the background gravitational waves with frequency $\nu_g\approx 10^{-17}$Hz. 
 
Recently, a body of works has devoted interest in the dynamics of dark energy by the emission of gravitational waves\cite{Romano, Singh}. This line of investigation is important since it can contribute to the innovation of studies involving dark energy, its cosmological role, phase of inflation and the relationship with the cosmological constant. 

\section{The power of emission of the black body associated to the dark sphere with Hubble radius}
 
According to the Planck law of the black body, the power of emission of radiation of a black body is given, namely: 
 
 \begin {equation}
  P=\frac{2\pi h A}{c^2}\int_0 ^\infty
  \frac{\nu^3}{e^{h\nu/K_B T}-1}d\nu, 
 \end {equation}
where $A$ is the area of the surface of the black body associated with the own surface of the dark sphere, i.e., $A=4\pi R_H^2$. So, we rewrite Eq.(8), as follows:

\begin {equation}
  P=\frac{8\pi^2 h R_H^2}{c^2}\int_0 ^\infty
  \frac{\nu^3}{e^{h\nu/K_B T}-1}d\nu.  
 \end {equation}
 
The emission power $P$ depends on $R_H$. In order to obtain also such a dependency in the distribution factor $(e^{h\nu/K_BT}-1)^{-1}$, let us write Eq.(9) in its equivalent form based on the toy model of dark energy sphere, since we are applying the idea of a gravitational analogue of black body to such a dark spherical surface ($2D$) representing the fabric of the space-time in this model of dimensional compactation. 
 
So, in doing this, let us first begin from Eq.(1), from where we obtain  
 
 \begin{equation}
 TR_H^2=\frac{G\hbar M_{dark}}{2\pi c K_B}, 
 \end{equation}
 from where we get 
 
 \begin{equation}
  K_B T=\frac{GhM_{dark}}{4\pi^2 c R_H^2},
 \end{equation}
where $\hbar=h/2\pi$ and $M=M_{dark}$.

By using Eq.(11), we can write the ratio $h\nu/K_BT$, which appears in Planck's law for black body in its equivalent form for representing the dark energy sphere, namely:

\begin{equation}
\frac{h\nu}{K_BT}=\frac{4\pi^2cR_H^2\nu}{GM_{dark}}. 
\end{equation}

Thus, according to Eq.(12), the Planck distribution factor applied to the dark sphere with radius $R_H$ is as follows: 

\begin{equation}
 \frac{1}{e^{(h\nu/K_BT)}-1}=
 \frac{1}{e^{(4\pi^2cR_H^2\nu/GM_{dark})}-1}, 
\end{equation}
where we can now see that the Planck distribution factor depends on the radius $R_H$ of the dark sphere.

Finally, substituting the equivalent distribution given by Eq.(13) in Eq.(9), we obtain 

\begin {equation}
P(R_H)=\frac{8\pi^2 h R_H^2}{c^2}\int_0 ^\infty
\frac{\nu^3}{e^{(4\pi^2cR_H^2\nu/GM_{dark})}-1}d\nu.  
 \end {equation}
 
Let us make the computational resolution of the integral in Eq.(14). So we write it as follows:

\begin {equation}
P(R_H)=\beta R_H^2\int_0 ^\infty
\frac{\nu^3}{e^{\alpha R_H^2\nu}-1}d\nu, 
 \end {equation}
where $\beta=8\pi^2h/c^2$ and 
$\alpha=4\pi^2c/GM_{dark}$, being $\beta\sim 10^{-49}Js^3/m^2$ and 
$\alpha\sim 10^{-25}s/m^2$. 

So, based on computational calculation, the power $P$ obtained for $r=R_H(\sim 10^{26}m)$ in the integral given in Eq.(15) is $P(R_H)\sim 10^{-73}W$. This power of emission of the dark energy of the whole universe is extremely low, being a very weak gravitational signal. 

Dark energy does not emit electromagnetic waves. However, it is important to call attention to the fact that such radiation with power in the order of $10^{-73}W$ is of gravitational nature (gravitational wave power).  Its frequency is extremely low in the order of $10^{-17}Hz$, which represents an isotropic background radiation of gravitational origin due to the accelerated expansion of the space-time. This means that when the fabric of space-time undergoes a stretching, background gravitational waves emerge on such dark fabric, i.e., the dark energy emits gravitational waves. 

It was shown that the cosmological constant is too small according to the observations, i.e., $\Lambda_0\sim 10^{-35}s^{-2}$ (or $\sim 10^{-52}m^{-2}$), being related to a small acceleration, namely $a_0\sim-10^{-9}m/s$. So, it is naturally expected that the frequency of the background gravitational wave is too low about $10^{-17}$Hz and with a wavelength of the order of $10^{25}$m. This means that such too low frequency has cosmological origin, thus being the lowest gravitational frequency within the wide range of frequencies of gravitational waves, i.e., $10^{-17}Hz<\nu<10^{3}Hz$\cite{Rowan}. This frequency of the order of $10^{-17}$Hz obtained by the present model is in agreement with data in the literature\cite{Rowan}. 

It is well-known that gravitational wave signals are expected over a wide range of frequencies, from $10^{-17}Hz$\cite{Rowan} in the case of ripples in the cosmological background to $10^3$Hz from the formation of neutron stars in supernova explosions. So, we notice that the present simple model was able to reproduce the order of magnitude of the lowest gravitational frequency based on the Unruh effect ($\nu\approx a_0/c\sim 10^{-17}Hz$) applied to the dark sphere (Fig.1) with a proof particle on its surface presenting a negative acceleration $a_0\sim-10^{-9}m/s^2$.  

\section{The cosmological scalar field $\Lambda(r)$ as alternative to the cosmological constant $\Lambda_0$ for explaining the background gravitational waves that control the dynamics of expansion of the universe} 

In a paper entitled {\it ``Cosmology on a Gravitational Wave Background''} (GWB)\cite{GWB}, where gravitational waves can explain the dynamics of the acceleration of the universe very well, a new model was developed to explain the current accelerating expansion of the universe where a GWB was incorporated by extending Einstein's equations to $R_{\mu\nu}-(1/2)Rg_{\mu\nu}+(2\pi^2/\lambda^2)g_{\mu\nu}=kT_{\mu\nu}$\cite{GWB}, where
$\lambda$ is the Compton wavelength of the graviton for the background gravitational wave. 

In this paper, the authors follow the idea of the reference\cite{GWB1}, where the space-time fluctuations produced by the big-bang are incorporated into Einstein`s equations, which 
contain an extra term $2\pi^2/\lambda^2$ that replaces the cosmological term by incorporating the energy of these fluctuations in space-time if $\lambda$ is the Compton wavelength of the graviton\cite{GWB}. 

In reference\cite{GWB1}, it was found that these fluctuations can explain the accelerated expansion of the universe and in reference\cite{GMB2}, it was shown that these fluctuations represented in this new term in Einstein’s equations are in agreement with the main observations of cosmology profiles of MPS and CMB. 

Here we intend to make a connection with the present toy model with the idea of the 
extra term $(2\pi2/\lambda^2)g_{\mu\nu}$\cite{GWB} that replaces the cosmological term 
$\Lambda g_{\mu\nu}$. Such extra term plays the role of the accelarating expansion of the 
universe without considering a cosmological constant that does not vary with the age of
the universe. 

Actually, in the present toy model we have shown that $\Lambda$ is in fact a variable
cosmological parameter $\Lambda(r)$, since it depends on the radius and time of the universe, although it reproduces the the well-known cosmological constant for $r=R_H$. This
is the reason why we have denominated $\Lambda(r)$ as a cosmological scalar field, which 
presents similarity with the term $2\pi^2/\lambda$\cite{GWB}, as we will show soon. 

According to Unruh effect, we have already obtained 

\begin{equation}
 \nu=\frac{K_BT}{h}=\frac{a}{4\pi^2 c}=\frac{G M_{dark}}{4\pi^2cR_H^2}, 
\end{equation}
where $a=GM_{dark}/R_H^2$. 

On the other hand, we write the cosmological scalar field $\Lambda(r)$ in the natural unity system, namely: 

\begin{equation}
\Lambda(r)=\frac{1}{c^2}\left[\frac{da}{dr}\right]=\frac{2GM_{dark}}{c^2r^3}
 \end{equation}
 where $\Lambda_0=\Lambda(r)|_{r=R_H}\sim 10^{-52} m^{-2}$ (cosmological constant), 
 where $M_{dark}\sim 10^{53}$Kg. 
 
Therefore, we reinforce the idea that the cosmological constant simply does not explain 
the entire dynamics of accelerated cosmic expansion given by the cosmological scalar field  $\Lambda(r)$ in Eq.(17). In this sense, the cosmological constant model must be abandoned.

Alternatively, we can write Eq.(17) in the following way:

\begin{equation}
 \Lambda(r)= \frac{2a}{c^2r}, 
\end{equation}
where $a=4\pi^2c\nu$ according to Eq.(7). So by substituting Eq.(7) in Eq.(18), we find

\begin{equation}
\Lambda(r)=\frac{8\pi^2\nu}{cr}, 
\end{equation}
where $\nu$ is the frequency of the graviton of the background gravitational wave. 

As $\nu=c/\lambda$, where $\lambda$ is the wavelength of the graviton of the 
background gravitational wave that controls the accelerated expansion of the universe
and plays the role of $\Lambda(r)$, replacing the invariant cosmological constant 
$\Lambda_O$, we write 

\begin{equation}
\Lambda(r)=\frac{8\pi^2}{\lambda r}, 
\end{equation}
so that the present toy model leads to the Einstein equation that provides 
the accelerated expansion of the universe, being consistent and very similar to the model of Cosmology on a Gravitational Wave Background (GWB)\cite{GWB}, namely: 

\begin{equation}
 R_{\mu\nu}-\frac{1}{2}Rg_{\mu\nu}+\left(\frac{8\pi^2}{\lambda r}\right)g_{\mu\nu}=
 kT_{\mu\nu}, 
\end{equation}
where $(8\pi^2/\lambda r)g_{\mu\nu}$ gives the accelerated expansion of the universe governed by the background gravitational wave. 

\section{Conclusions}  

In this paper, we were able to propose a toy model to represent the dark sector of the universe (dark energy) by means of a spherical shell of dark energy with Hubble radius $R_H$. Such a spherical surface behaves like a black body with a very low temperature (a gravitational analogue of black body), having a dark energy emitting a frequency of approximately $10^{-17}$Hz, which is in the same order of magnitude of the frequency of background gravitational waves as being a fossil of the inflationary period. Such frequency is emitted due to the acceleration of the cosmic expansion (expansion of the spherical surface representing the fabric of space-time like a balloon that grows up) generated by the cosmological constant whose small order of magnitude was also obtained by this model acoording to the observations. 

As we made an analogy of the dark sphere with a black body, we were also able to obtain the emission power of the dark energy sphere, which is a very low power by indicating that the gravitational wave signal emitted by the fabric of space-time is extremely weak. So, this would make it very difficult to detect.

Although the present model has used the Newtonian approximation for obtaining
Eq.(3) in the current scenario of the universe, the early universe had a too large acceleration associated with a very large value of $\Lambda$, which leads to the cosmic inflation. This issue can be explored in the future, since recently it was discovered that in the exotic conditions of the early universe, gravitational waves may have shaken the space-time so hard that they spontaneously created electromagnetic radiation\cite{Sugumi}. 

Here it is important to stress that the model based on an expanding two-dimensional surface is related to the bulk ($3$-dimensional volume) relating and conserving all the corresponding dimensional physical properties (e.g: the universal gravitational law and anti-gravity connected to the gases properties by mean of the Unruh effect). This concept is explored in Brane models that predict extra dimensions in the Universe by describing physical properties and conservation laws that are not compromised or invalidated by the dimensional reduction of the Manifold. In other words, physical properties and laws remain unchanged when the manifold is dimensionally reduced. These foundations were substantially explored in works inspired by Cosmological Branas Models\cite{RS99,RS20}, which reinforces the validity of the dimensional compaction toy model we are using.


\begin{thebibliography}{99}

\bibitem{Maartens} R. Maartens,{\it J. Phys.: Conf. Ser.} {\bf 68}  012046 (2007).

\bibitem{Arkani} N. Arkani-Hamed, S. Dimopoulos, N. Kaloper and R. Sundrum, {\it Phys.Lett. B} {\bf 480}  193 (2000).

\bibitem{Langlois} D. Langlois, M. Rodrigues-Martinez, {\it Phys. Rev. D} {\bf 64}  123507 (2001). 
 
\bibitem{Goldberger} W. D. Goldberger and M. B. Wise, {\it Phys. Rev. D} {\bf 60}  107505 (1999).

\bibitem{DeWolfe} O. DeWolfe, D. Z. Freedman, S. S. Gubser and A. Karch, 
{\it Phys.Rev.D} {\bf 62} 046008 (2000).

\bibitem{Bazeia} D. Bazeia, C. Furtado and A. R. Gomes,
{\it JCAP} {\bf 2004} {\bf N.2} 002 (2004).

\bibitem{1} A. D. Sakharov, {\it Sov. Phys. Dokl.} {\bf 12}  1040 (1968). 

\bibitem{2} T. Jacobson, {\it Phys. Rev. Lett.} {\bf 75}  1260 (1995). 

\bibitem{3} G. E. Volovik, {\it Phys. Rep.} {\bf 351}  195 (2001). 

\bibitem{4} M. Visser, {\it Mod. Phys. Lett. A} {\bf 17}  977 (2002). 

\bibitem{5} C. Barcelo et al, {\it Int. J. Mod. Phys. D} {\bf 10}  799 (2001). 

\bibitem{6} G. E. Volovik, {\it The Universe in a Helium Droplet}, 
 {\it Oxford Uni. Pres} (2003). 
 
\bibitem{7} G. Jannes, arXiv:0907.2839. 

\bibitem{8} H. S. Yang and M. Sivakumar, arXiv:0908.2809. 

\bibitem{9} J. Makela, arXiv:0805.3952. 

\bibitem{10} E. Elizalde and J. S. Pedro, {\it Phys. Rev. D} {\bf 78}  061501 (2008). 

\bibitem{11} C. J. Hogan, {\it Phys. Rev. D} {\bf 77}  104031 (2008).

\bibitem{12} K. Bamba et al, arXiv:0909.4397. 

\bibitem{13} G. Chirco and S. Liberati, arXiv:0909.4194. 

\bibitem{14} L. Sindoni, F. Girelli and S. Liberati, arXiv:0909.5391. 

\bibitem{15} K. Bamba, C. Cheng and S. Tsujikawa, arXiv:0909.2159. 

\bibitem{16} C. Barcelo, S. Liberati and M. Visser, arXiv:0505065. 

\bibitem{17} T. Padmanabhan, {\it Class. Quantum. Grav.} {\bf 19}  5387 (2002). 

\bibitem{18} T. Padmanabhan, {\it Phys. Rep.} {\bf 406}  49 (2005). 

\bibitem{19} T. Padmanabhan, {\it Gen. Rel. Grav.} {\bf 40}  2031 (2008).

\bibitem{Lambda} C. Nassif and A. C. Amaro de Faria Jr, 
{\it Reviews in Physics} {\bf 11}  100088 (2023).

\bibitem{20} W. G. Unruh, {\it Physics meets Philosophy at the Planck Scale},
{\bf N.4}  152 (2001).

\bibitem{21} S. M. Carroll, arXiv:0107571. 

\bibitem{22} M. Carmeli and T.Kuzmenko, arXiv:0102033.

\bibitem{23} B. P. Schmidt,et al., {\it Astrophys. J.} {\bf 507}  46 (1998).

\bibitem{24} A. G. Riess, et al., {\it Astronom. J.} {\bf 116}  1009 (1998).

\bibitem{25} P. M. Garnavich, et al., {\it Astrophys. J.} {\bf 509} 74 (1998).

\bibitem{26} S. Perlmutter, et al., {\it Astrophys. J.} {\bf 517}  565 (1999).
\bibitem{27} M. S. Turner and L. M. Krauss, {\it Gen. Rel. Grav.}  {\bf 27}    1137 (1995).  

\bibitem{28} Y. Wang and M. Tegmark, {\it Phys. Rev. Lett.} {\bf 92} 241302 (2004).

\bibitem{29} C. Nassif, {\it Int. Journal of Modern Physics D} {\bf 25},
{\bf N.10} 1650096 (2016).  

\bibitem{30} Ya.B. Zeldovich, {\it JETP Lett.} {\bf 6} 316 (1967);
S. Weinberg, {\it Rev. Mod. Phys.} {\bf 61}  1 (1989);
T. Padmanabhan, {\it Phys. Rept.} {\bf 380}  235 (2003). 

\bibitem{Romano} A. E. Romano, arXiv:2211.05760. 

\bibitem{Singh} A. Singh, {\it Phys. Letters B} {\bf 802}  135226 (2022).

\bibitem{Rowan} S. Rowan and J. Hough, {\it The detection of gravitational waves},\\
https://cds.cern.ch/record/454173/files/open-2000-258.pdf. 

\bibitem{Sugumi} S. Kanno, J. Soda and K. Ueda, 
{\it Physical Review D}  {\bf 106} 083508 (2022). 

\bibitem{RS99} L. Randall and R. Sundrum, {\it Phys. Rev. Lett.} {\bf 83}   3370 (1999). 

\bibitem{RS20} L. Randall and R. Sundrum, {\it Phys. Rev. Lett.} {\bf 83} 
4690 (1999).

\bibitem{GWB} T. Matos, L. A. Escamilla, M. Hernández, J. A. Vázquez, 
{\it MNRAS}  {\bf 529}  3013 (2024). 

\bibitem{GWB1} T. Matos and L. L. Parrilla, {\it Rev. Mex. Fis.} {\bf 67},
{\bf N.4} 040703 (2021). 

\bibitem{GMB2} T. Matos and L. O. Tellez-Tovar, {\it Rev. Mex. Fis.} 
 {\bf 68}, {\bf N.2}  020705  (2022).
\end{thebibliography}
\end{document}